# Cryptanalysis of a Public-key Cryptosystem Using Lattice Basis Reduction Algorithm

Roohallah Rastaghi [1], Hamid R. Dalili Oskouei [2]

[1,2] Department of Electrical Engineering, Aeronautical University of Since & Technology,
Tehran, Iran

**Abstract**
In this paper, we proposed a new attack against Hwang et *al*.'s cryptosystem. This cryptosystem uses a super-increasing sequence as private key and the authors investigate a new algorithm called permutation combination algorithm to enhance density of knapsack to avoid the low-density attack. Sattar J. Aboud [Aboud j. Sattar, "An improved knapsack public key cryptography system", International Journal of Internet Technology and Secured Transactions, Vol.3 (3), pp.310-319, 2011] used Shamir's attack on the basic Merkle-Hellman cryptosystem to break this cryptosystem. Due to use of Lenstera's integer programming, Lagarias showed that Shamir's attack is inefficient in practice; So, Aboud's attack is impractical too.

In this paper, we introduce a direct attack against Hwang et *al*.'s cryptosystem based on Lattice basis reduction algorithms. By computing complexity of propose attack, we show that unlike Aboud's cryptanalysis, our cryptanalysis is more efficient and practicable.

## 1. Introduction

The first knapsack-type public key cryptosystem (PKC) was introduced by Merkle and Hellman[12]. Since its proposal, knapsack-type PKC had been widely studied and many knapsack PKCs were developed. However, almost all knapsack cryptosystems were shown insecure in that they are vulnerable to some known attacks, such as low density attack [2,8], orthogonal lattice attack [15], ... .

Nowadays, we reconsider knapsack public key cryptography because Shor [17] showed that integer factorization and discrete logarithm problems can be easily solved by using quantum computers. Therefore, traditional public key cryptosystem based on the two problems cannot be used to provide privacy protections any longer and public key cryptosystems secure in quantum computing environments are needed to be developed. The knapsack problem is NP-complete [14]. Hence, we can design cryptosystems based on the knapsack problem in order to resist quantum attacks. On the other hand, although the underlying problem is NP-complete, but some of the knapsack cryptosystems such as Merkle-Hellman [16], Chor-Rivest [18], was broken due to the special structure of the private key and the mathematical way that public key (public knapsack) was built from the private key.

M. S. Hwang et *al*. [6] introduced a new knapsack type public key cryptosystem in 2009. This cryptosystem is based on basic Merkel-Hellman knapsack cryptosystem [12] and uses a super-increasing sequence as private key. They investigate a new algorithm called permutation combination algorithm. By exploiting this algorithm, the authors attempted to enhance density of knapsack to avoid the low-density attack. Hwang et *al*. knapsack-type cryptosystem was attacked by Aboud [1]. Aboud's attack is based on Shamir's attack [16] on the basic Merkle-Hellman cryptosystem. Lagarias [9] showed due to use of Lenstera's integer programming, Shamir's attack is inefficient in practice, so, Aboud's attack is not practicable.

In this paper, we use LLL-lattice basis reduction algorithm for analysis Hwang et *al*.'s knapsack-type cryptosystem. The LLL-lattice basis reduction algorithm is a crucial component in many number-theoretic algorithms. It is useful for solving certain knapsack (subset sum) problems, and has been used for cryptanalyzeing public-key encryption schemes which are based on the subset sum problem. We show that because of the special structure in the key generation stage, we can use the LLL-lattice basis reduction algorithm for cryptanalyzeing Hwang et *al*.'s cryptosystem and obtain equivalent private keys (super-increasing sequences).

The rest of this paper is organized as follows: In the following section, we briefly explain some mathematical background. These concepts are useful for understanding the security analysis of the Hwang et *al*.'s cryptosystem. Then, in Section 3, we review Hwang et *al*.'s knapsack cryptosystem. New cryptanalysis of this cryptosystem will be discussed in Section 4 and in section 5, we compute the computational complexity of the proposed attack.

## 2. Mathematical Background

In this section, we recall some concepts about the subset-sum problem and lattice theory. These concepts are useful to understand the security analysis of the Hwang et *al.*'s cryptosystem.

The subset sum problem is stated as follows:

**Definition1.** (Subset sum problem (SSP)). *A set of positive integers* $(a_1, a_2, \ldots, a_n)$ *and positive integer s is given. Whether there is a subset of the $a_i$'s that their sum equal to s. That is equivalent to determine whether there are variables* $x_i \in \{0,1\}, 1 \leq i \leq n$ *such that*

$$\sum_{i=1}^{n} x_i \, a_i = s$$

The Subset sum problem is a particular case of the 0-1 knapsack problem. The subset sum problem has been proven to be *NP-complete*. The computational version of the subset sum problem is *NP-hard* [13].

**Definition2.** (super-increasing sequence). ) *The sequence* $(a_1, a_2, \ldots, a_n)$ *of positive integers is a super-increasing sequence, if* $a_i > \sum_{j=1}^{i-1} a_j$ *for all* $i \geq 2$.

There is an efficient greedy algorithm to solve the subset sum problem if the $a_i$ are a super-increasing sequence: Just subtract the largest possible value from $s$ and repeat. Algorithm1 efficiently solves the subset sum problem for super-increasing sequences in the polynomial time.

---
**Algorithm1:** Solving a super-increasing subset sum problem

---
INPUT: super-increasing sequence $(a_1, a_2, \ldots, a_n)$ and an integer $s$ which is the sum of a subset of the $a_i$ s.
OUTPUT: $(x_1, x_2, \ldots, x_n)$ where $x_i \in \{0,1\}$, such that $s = \sum_{i=1}^{n} a_i . x_i$.

1. $i \leftarrow n$
2. While $i \geq 1$ do the following:
2.2. If $s \geq a_i$ then $x_i \leftarrow 1$ and $s \leftarrow s - a_i$ Otherwise $x_i \leftarrow 0$.
2.3. $i \leftarrow i - 1$
3. Return( $(x_1, x_2, \ldots, x_n)$ ).

---

Knapsack public-key encryption schemes are based on the subset sum problem, which is NP-complete. The basic idea is to select an instance of the subset sum problem that is easy to solve, and then to disguise it as an instance of the general subset sum problem which is hopefully difficult to solve. The original knapsack set can serve as the private key, while the transformed knapsack set serves as the public key.

**Definition3.** *Let* $x = (x_1, x_2, \ldots, x_n)$ *and* $y = (y_1, y_2, \ldots, y_n)$ *be two vectors in* $\mathbb{R}^n$. *The inner product of x and y is the real number*

$$<x, y> = x_1 y_1 + x_2 y_n + \ldots + x_n y_n$$

**Definition4.** *Let* $y = (y_1, y_2, \ldots, y_n)$ *be a vector in* $\mathbb{R}^n$. *The $\ell_2$ norm (or Euclid norm) of y is the real number*

$$\|y\| = \sqrt{<y, y>} = \sqrt{y_1^2 + y_2^2 + \cdots + y_n^2}.$$

The *sup norm, maximum norm or* $\ell_\infty$ norm is:

$$\|y\|_\infty = \max_{i=1,2,\ldots,n} |y_i|$$

We can show that

$$\|y\|_\infty \leq \|y\|$$

**Definition5.** *Let* $\{f_1, f_2, \ldots, f_m\}$ *be a set of linearly independent vectors in* $\mathbb{R}^n$ *(m $\leq$ n). The set L of all integer linear combinations of* $f_1, f_2, \ldots, f_m$ *is called a lattice of dimension m; that is*

$$L = \left\{ \sum_{i=1}^{m} l_i f_i \; : \; l_i \in \mathbb{Z} \right\}.$$

*The vectors* $f_1, f_2, \ldots, f_m$ *are called a basis for the lattice L.*

**Definition6.** (Gram-Schmidt orthogonalization).

*Let* $\{f_1, f_2, \ldots, f_n\}$ *be an arbitrary basis of* $\mathbb{R}^n$. *Define the vectors* $f_i^*, 1 \leq i \leq n$ *inductively by*

$$f_i^* = f_i - \sum_{j=1}^{i-1} \mu_{i,j} f_j^*$$

*Where* $\mu_{i,j} = \frac{<f_i, f_j^*>}{<f_j^*, f_j^*>}$ *for* $1 \leq i \leq n$. *In particular* $f_1^* = f_1$. *We will call* $f_1^*, f_2^*, \ldots, f_n^*$ *the Gram-Schmidt orthogonal basis of* $\{f_1, f_2, \ldots, f_n\}$ *and the* $f_i^*$ *together* $\mu_{i,j}$ *form the Gram-Schmidt orthogonalization of* $\{f_1, f_2, \ldots, f_n\}$.

A lattice can have many different bases. A basis consisting of vectors of relatively small lengths is called *reduced*. The following definition provides a useful notion of a reduced basis, and is based on the Gram-Schmidt orthogonalization procedure.

**Definition 7.** Let $f_1, f_2, \ldots, f_n \in \mathbb{R}^n$ be linearly independent and $f_1^*, f_2^*, \ldots, f_n^*$ the corresponding Gram-Schmidt orthogonal basis. Then $\{f_1, f_2, \ldots, f_n\}$ is reduced if $\|f_i^*\| \leq 2\|f_{i+1}^*\|$ for $1 \leq i < n$. So we have

$$\|f_1\|^2 = \|f_1^*\|^2 \leq 2\|f_2^*\|^2 \leq 2^2\|f_3^*\|^2 \leq \cdots \leq 2^{n-1}\|f_n^*\|^2.$$

The basis $\{f_1, f_2, \ldots, f_n\}$ is said to be *reduced* (more precisely, *Lovász-reduced*) if

$$|\mu_{i,j}| \leq \frac{1}{2}, \quad for \ 1 \leq j < i \leq n$$

(where $|\mu_{i,j}|$ denotes the absolute value of $\mu_{i,j}$), and

$$\|f_i^*\|^2 \geq \left(\frac{3}{4} - \mu_{i,i-1}^2\right)\|f_{i-1}^*\|^2 \quad for \ 1 < i < n.$$

The LLL-lattice basis reduction algorithm is a crucial component in many number-theoretic algorithms such as simultaneous Diophantine approximation Problem. It is useful for solving certain subset sum problems, and has been used for cryptanalyzeing public-key encryption schemes which are based on the subset sum problem.

**Algorithm 2**[13]: LLL-lattice basis reduction algorithm

INPUT: a basis $\{f_1, f_2, \ldots, f_n\}$ for a lattice $L$ in $\mathbb{R}^n, m \geq n$.
OUTPUT: a reduced basis for $L$.
1. $f_1^* \leftarrow f_1$, $F_1 \leftarrow <f_1^*, f_1^*>$.
2. For $i$ from 2 to $n$ do the following:
2.1  $f_i^* \leftarrow f_i$.
2.2  For $j$ from 1 to $i-1$, set $\mu_{i,j} \leftarrow <f_i, f_j^*>/F_j$ and $f_i^* \leftarrow f_i^* - \mu_{i,j} f_j^*$.
2.3  $F_i \leftarrow <f_i^*, f_i^*>$.
3. $k \leftarrow 2$.
4. Execute subroutine $RED(k, k-1)$ to possibly update some $\mu_{i,j}$.
5. If $F_k < (\frac{3}{4} - \mu_{k,k-1}^2)F_{k-1}$ then do the following:
5.1 Set $\mu \leftarrow \mu_{k,k-1}$, $F \leftarrow F_k + \mu^2 F_{k-1}$,
     $\mu_{k,k-1} \leftarrow \mu F_{k-1}/F$, $F_k \leftarrow F_{k-1}F_k/F$ and
     $F_{k-1} \leftarrow F$.
5.2 Exchange $f_k$ and $f_{k-1}$.
5.3 If $k > 2$ then exchange $\mu_{k,j}$ and $\mu_{k-1,j}$ for
     $j = 1, 2, \ldots, k-2$.
5.4 For $i = k+1, k+2, \ldots, n$:
     Set $t \leftarrow \mu_{i,j}, \mu_{i,k} \leftarrow \mu_{i,k-1} - \mu t$, and
     $\mu_{i,k-1} \leftarrow t + \mu_{k,k-1} - \mu_{i,k}$.
5.5 $k \leftarrow \max(2, k-1)$.
5.6 Go to step 4.
     Otherwise, for $l = k-2, k-3, \ldots, 1$, execute
     $RED(k, l)$, and finally set $k \leftarrow k+1$.
6. If $k \leq n$ then go to step 4. Otherwise, return $\{f_1, f_2, \ldots, f_n\}$.

$RED(k, l)$: If $|\mu_{k,l}| > 1/2$ then do the following:
1.   $r \leftarrow \lfloor 0.5 + \mu_{k,l} \rfloor, f_k \leftarrow f_k - rf_l$.
2.   For $j$ from 1 to $l-1$, set $\mu_{k,j} \leftarrow \mu_{k,j} - r\mu_{l,j}$.
3.   $\mu_{k,l} \leftarrow \mu_{k,l} - r$.

The LLL-lattice basis reduction algorithm is a polynomial-time algorithm for finding a reduced basis, given a basis for a lattice.

**Theorem 1.** Let $L \subset \mathbb{Z}^n$ be a lattice with basis $\{f_1, f_2, \ldots, f_n\}$, and let $C \in \mathbb{R}$, $C \geq 2$ be such that $\|f_i\|^2 \leq C$ for $i = 1, 2, \ldots, n$. Then the number of arithmetic operations needed by Algorithm 2 is $O(n^4 \log C)$, on integers of size $O(n \log C)$ bits, which is polynomial time.

**Proof:** see [11].

**Lemma 1.** Let $f_1, f_2, \ldots, f_n$ be a LLL reduced basis of a rational lattice $L \subset \mathbb{Q}^n$ and $f_1^*, f_2^*, \ldots, f_n^*$ be its Gram-Schmidt orthogonalization. Then

$$det(L) = \prod_{i=1}^{n} |f_i^*|.$$

**Proof:** see [5].

**Definition 8.** (Simultaneous Diophantine Approximation Problem). Let $a_1, a_2, \ldots, a_n \in \mathbb{R}$ and let $\epsilon > 0$. Let $Q \in \mathbb{N}$ be an integer such that $Q \geq \epsilon^{-n}$. The simultaneous diophantine approximation problem is to find $(q, p_1, \ldots, p_n) \in \mathbb{Z}$ such that $0 < q \leq Q$ and

$$|a_i - p_i/q| \leq \epsilon/q$$

for all $1 \leq i \leq n$.

A major application of algorithm 2 is to give an algorithm to compute the integers $(q, p_1, \ldots, p_n)$ in Definition 8. In practice, the real numbers $a_1, a_2, \ldots, a_n$ are given to some decimal precision (and so are rational numbers with coefficients of some size). The size of an instance of the simultaneous Diophantine approximation is the sum of the bit lengths of the numerator and denominator of the given approximations to the $a_i$, together with the bit length of the representation of $\varepsilon$ and $Q$. Let $X$ be a bound on the absolute value of all numerators and denominators of the $a_i$. The computational task is to find a solution $(q, p_1, \ldots, p_n)$ in time which is polynomial in $n$, $\log(X)$, $\log(1/\varepsilon)$ and $\log Q$.

**Theorem2.** (Solving the simultaneous Diophantine approximation problem). *Let $a_1, a_2, ..., a_n \in \mathbb{Q}$ be given as rational numbers with numerator and denominator bounded in absolute value by X. Let $0 < \epsilon < 1$. One can compute in polynomial time integers $(q, p_1, ..., p_n)$ such that $0 < q < 2^{n(n+1)/4} \epsilon^{-(n+1)}$ and $|a_i - p_i/q| \leq \epsilon/q$ for all $1 \leq i \leq n$.*

**Proof:** A general proof of this theorem is given in [7] but we introduce different and simple proof.
Let $Q = 2^{n(n+1)/4} \epsilon^{-n}$ and $L \subseteq \mathbb{Q}^{n+1}$ be the lattice by the rows $f_0, ..., f_n \in \mathbb{Q}^{n+1}$ of the matrix

$$\begin{pmatrix} \epsilon/Q & a_1 & a_2 & \cdots & a_n \\ 0 & -1 & 0 & \cdots & 0 \\ 0 & 0 & -1 & & 0 \\ \vdots & \vdots & & \ddots & \vdots \\ 0 & 0 & & \cdots & -1 \end{pmatrix}$$

The dimension is $n+1$ and the determinant is $\epsilon/Q = 2^{-n(n+1)/4} \epsilon^{n+1}$. The entries of the lattice are ratios of integers with absolute value bounded by $max\{X, \frac{2^{n(n+1)/4}}{\epsilon^{n+1}}\}$. Note that the lattice $L$ does not have a basis with entries in $\mathbb{Z}$, but rather in $\mathbb{Q}$.

The LLL algorithm applied to $L$, outputs a non-zero vector
$v = (v_1, ..., v_n) = (q\epsilon/Q, qa_1 - p_1, qa_2 - p_2, ..., qa_n - p_n)$.
If $v$ is the smallest vector found by the LLL-algorithm, then from definition7, we have

$$\|v\| \leq \|f_0^*\|^2 \leq 2\|f_1^*\|^2 \leq 2^2\|f_2^*\|^2 \leq \cdots \leq 2^n\|f_n^*\|^2.$$

By multiplying together the $n+1$ above inequality, we have

$$\|v\|^{2(n+1)} \leq \|f_0^*\|^2 . 2\|f_1^*\|^2 ... 2^n\|f_n^*\|^2$$
$$\leq 2^{n(n+1)/2} \|f_0^*\|^2 ... \|f_n^*\|^2$$

Hence,
$$\|v\| \leq 2^{n/4}(\|f_0^*\| ... \|f_n^*\|)^{1/(n+1)}$$

From lemma 1, we know that $det(L) = \prod_{i=1}^{n} |f_i^*|$ and so

$$\|v\| \leq 2^{n/4} det(L)^{1/(n+1)} = 2^{n/4} 2^{-n/4} \epsilon = \epsilon < 1.$$

If $q = 0$ then $v = (0, -p_1, -p_2, ..., -p_n)$ with some $p_i \neq 0$ and so $\|v\| \geq 1$, hence $q \neq 0$. Without loss of generality, $q > 0$. Since $\|v\|_\infty \leq \|v\|$ and $\|v\| < \epsilon < 1$ it follows that $q\epsilon/Q < \epsilon < 1$ (where $\|v\|_\infty = q\epsilon/Q = v_1$) and so

$$0 < q < Q/\epsilon = 2^{n(n+1)/4} \epsilon^{-(n+1)}.$$

Similarly, for other $v_i$, $2 \leq i \leq n$ we have $|qa_i - p_i| \leq \epsilon$ and so

$$|a_i - p_i/q| \leq \epsilon/q \quad \text{for } 1 \leq i \leq n.$$

**Theorem3.** *If we use LLL algorithm for solving Simultaneous Diophantine Approximation Problem, then the computational complexity of the problem is $O(n^6 \max\{n \log(X), n^2 + n \log(1/\epsilon)\}^3)$, which is polynomial time.*
**Proof:** See [7].

## 3. Hwang Et Al.'s Cryptosystem

Hwang *et al.*'s cryptosystem is based on the basic Merkle-Hellman knapsack cryptosystem.

3.1 Key Generation:

Each user chooses a super-increasing sequence $(b_1, b_2, ..., b_{1360})$ as secret key. i.e.

$$b_i > \sum_{j=1}^{i-1} b_j \quad (i = 1, 2, ..., 1360).$$

Choose a large prime $p$ as modulus such that $p > \sum_{i=1}^{1360} b_i$, two modular multipliers $w$ and $w'$ such that $gcd(p, w) = 1$, and $w.w' = 1 \mod p$. Each user transfers super-increasing sequence $B = (b_1, b_2, ..., b_{1360})$ into a pseudorandom sequence $A = (a_1, a_2, ..., a_{1360})$ as follows:

$$a_i = b_i.w \mod p, \quad (1 \leq i \leq 1360) \quad (1)$$

The public key is $(a_1, a_2, ..., a_{1360})$ and the private key is $\{(b_1, b_2, ..., b_{1360}), w, w', p\}$.

They presented a permutation combination algorithm and used this algorithm to ensure the security of the cryptosystem. By exploiting this algorithm, they attempted to enhance density of knapsack to avoid the low-density attack [2, 8]. The permutation algorithm is as follows:

1. Define an original sequence
$$D_0 = \{E_n, E_{n-1}, E_{n-2}, ..., E_5, E_4, E_3, E_2, E_1\}.$$

2. Recombine all the elements of the original sequence $D_0$ which obtain $(n! - 1)$ sequences $D_1, ..., D_{(n!-1)}$. The sequences $D_i (i = 1, 2, ..., n! - 1)$ are defined as follows:
$$D_0 = \{E_n, E_{n-1}, E_{n-2}, ..., E_5, E_4, E_3, E_2, E_1\}$$
$$D_1 = \{E_n, E_{n-1}, E_{n-2}, ..., E_5, E_4, E_3, E_1, E_2\}$$
$$\vdots$$
$$D_{n!-1} = \{E_1, E_2, E_3 E_4, ..., E_{n-2}, E_{n-1}, E_n\}$$

3. Suppose we can compute $D_m$ for $1 \leq m \leq n! - 1$. $m$ can be written as
$$m = \sum_{i=1}^{n} u_i(n-i)!, \quad 0 \leq u_i \leq n.$$

Each sequence has an own corresponding value called the

*factorial carry value*, $\{u_n, u_{n-1}, \ldots, u_2, u_1\}$. Using the factorial carry value, we can efficiently obtain any sequence with the following algorithm.

**Algorithm 3:** permutation combination algorithm

INPUT: $D_0 = (E_1, E_2, \ldots, E_n)$ and integers $m$.

OUTPUT: $D_m = (E'_1, E'_2, \ldots, E'_3)$.

$m = \sum_{i=1}^{n} u_i \times (n-i)!$.

For $1 \leq i \leq n$ do
    if $u_i = 0$ then
        $E'_i = E_i$ ;
    else {
        for $(1 \leq j \leq u_i)$ do
            $E'_{i+j} = E_i$ };

Return $(E'_1, E'_2, \ldots, E'_3)$.

For instance, generate the original vector $D_0 = (A, B, C, D, E, F)$. Find the result of $D_{100}$ : $100 = 0 \times 5! + 4 \times 4! + 0 \times 3! + 2 \times 2! + 0 \times 1! + 0$ then $D_{100} = (A, F, B, E, C, D)$.

### 3.2 Encryption:

For encrypt the message $M$, the sender executes the following steps:

1. Select a hash function whose digest is 1024 bits and compute the digest $D$ of $M$ as $D = H_{1024}(M)$.
2. Compute $D' = D \mod 170!$
3. Compute the factorial carry value $U = \{u_1, u_2, \ldots, u_{170}\}$ of $D'$ where $D' = u_1 \times 169! + u_2 \times 168! + \cdots + u_{170} \times 0!$
4. Divide the public key vector $A = (a_1, a_2, \ldots, a_{1360})$ into 8 subset public key vectors. Each subset public key vector has 170 elements.

$$A = \{(a_1, a_2, \ldots, a_{170}), \ldots, (a_{1191}, a_{1192}, \ldots, a_{1360})\}.$$

5. Recombine each subset public key vector using $U = \{u_1, u_2, \ldots, u_{170}\}$ by means of the permutation combination Algorithm. Then chooses the first 128 elements in each subset public key vector. Thus, the sender obtain 1024 elements $A_u = (au_1, au_2, \ldots, au_{1024})$.
6. The message $M$ is divided into $\{M_1, M_2, \ldots, M_j\}$. Each $M_k$ ($k = 1, 2, \ldots, j$) is a 1024-bit message:

$$M_k = \{x_{k,1}, \ldots, x_{k,1024}\}$$

7. The corresponding ciphertext $C_k$ is given as the product of $A_u = (au_1, au_2, \ldots, au_{1024})$ and $M_k$ ($k = 1, 2, \ldots, j$).

$$C_k = \sum_{i=1}^{1024} au_i \times x_{k,i}, \quad 1 \leq k \leq j.$$

The ciphertext is $C = \{C_1, \ldots, C_j\}$ and sends $(C, D')$ to the receiver.

### 3.3 Decryption:

Receiver after receiving $(C, D')$, executes the following steps to derive $M$ from $C$ and $D'$:

1. Compute the factorial carry value $U = \{u_1, u_2, \ldots, u_{170}\}$ of $D'$ where

$$D' = u_1 \times 169! + u_2 \times 168! + \cdots, u_{170} \times 0.$$

2. Divide his/her secret key vector $B = (b_{b1}, b_{b2}, \ldots, b_{b1360})$ into 8 subset public key vectors. Each key vector has 170 elements.

$$B = \{(b_1, b_2, \ldots, b_{170}), \ldots, (b_{1191}, b_{1192}, \ldots, b_{1360})\}.$$

3. Recombine each subset public key vector using $U = \{u_1, u_2, \ldots, u_{170}\}$ by means of the Permutation Combination Algorithm. Then chooses the first 128 elements in each subset public key vector. The receiver obtain 1024 elements $B_u = (bu_1, bu_2, \ldots, bu_{1024})$. However $B_u = (bu_1, bu_2, \ldots, bu_{1024})$ is still a super-increasing sequence.
5. Divide $C$ into $C = \{C_1, \ldots, C_j\}$. Each $C_k$ ($k = 1, 2, \ldots, j$) is a 1024-bit ciphertext.
6. Compute

$$D_k = C_k \times w' \mod p$$
$$= \sum_{i=1}^{1024} (au_i \times x_{k,i}) \times w' \mod p$$
$$= \sum_{i=1}^{1024} (bu_i \times w \times x_{k,i}) \times w' \mod p$$
$$= \sum_{i=1}^{1024} bu_i \times x_{k,i} \mod p$$

for $k = 1, 2, \ldots, j$. So we have $D_k = \sum_{i=1}^{1024} bu_i \times x_{k,i} \mod p$. Since $p > \sum_{i=1}^{1360} b_i$ we have $D_k = \sum_{i=1}^{1024} bu_i \times x_{k,i}$. Hence, the receiver can solves these super-increasing knapsack problems with algorithm1 and obtains $x_{k,i}$ for $1 \leq i \leq 1024$ and $1 \leq k \leq j$. Therefore, we can recover original message $M = \{M_1, M_2, \ldots, M_j\}$ where $M_k = \{x_{k,1}, \ldots, x_{k,1024}\}$.

Aboud attacked this cryptosystem by using Shamir's attack [16] on the basic Merkle-Hellman cryptosystem. As we said, Lagarias in [9] showed Shamir's attack is inefficient in practice, so Aboud's attack is not practicable.

# 4. Our Proposed Attack

In this section, we present our attack against the Hwang et al.'s knapsack cryptosystem. The first step in the attack is noticing that the given knapsack problem $c = \sum_{i=1}^{n} a_i m_i$ (with public weights $(a_1, a_2, \ldots, a_n)$ and target $c$) can be transformed into *infinitely many different easy knapsack problems with super-increasing weights* $(a'_1, a'_2, \ldots, a'_n)$ *and target $c'$*. This was independently observed by Eier-Lagger [4] and Desmedt-Vanderwalle-Govaerts [3]. Their result can be summarized in the following lemma.

Let $(b_1, b_2, \ldots, b_n)$ be the private super-increasing sequence, $(a_1, a_2, \ldots, a_n)$ be the Corresponding public key such that $a_i = w \cdot b_i \bmod p$ and $W, p$ be defined as in section 3. Let $U = w^{-1} \bmod p$, so we have

$$b_i = U \cdot a_i \bmod p.$$

**Lemma 2.** *There exists an $\varepsilon > 0$ such that if $\frac{U'}{p'}$ is rational with $\left|\frac{U'}{p'} - \frac{U}{p}\right| < \varepsilon$, then the weights $(b'_1, b'_2, \ldots, b'_n)$ where $b'_i = U'a_i \bmod p'$ for $i = 1, \ldots, n$ are super-increasing.*

Our attack consists of three steps: in step1, we can use LLL-lattice basis reduction algorithm for finding a super-increasing sequence $B' = (b'_1, b'_2, \ldots, b'_{1360})$ that is very close to super-increasing sequence $B = (b_1, b_2, \ldots, b_n)$. In step2 and step3 we use super-increasing sequence $B' = (b'_1, b'_2, \ldots, b'_{1360})$ and public ciphertext $(C, D')$ for recover the plaintext.

**Step1:**
In the general form, equation (1) can be written as follows:

$$a_i = b_i \cdot w \bmod p, \quad 1 \le i \le n.$$

Where $(a_1, a_2, \ldots, a_n)$ is the public key and $(b_1, b_2, \ldots, b_n)$ is the private key.

Let $U = w^{-1} \bmod p$ where $1 \le U < p$. We have

$$b_i = a_i \cdot w^{-1} \bmod p = a_i \cdot U \bmod p, 1 \le i \le n \quad (2)$$

This means that for $1 \le i \le n$, there exists some integers $k_i$ such that

$$a_i U - k_i p = b_i$$

and $0 \le k_i < a_i$. Hence,

$$0 \le U/p - k_i/a_i = b_i/a_i p. \quad (3)$$

Since $(b_1, b_2, \ldots, b_n)$ is a super-increasing sequence, so $b_i \ge 2^{i-1}$ and with $p > \sum_{i=1}^{n} b_i$ we have

$$0 \le b_i < p/2^{n-i}$$

Hence

$$0 \le U/p - k_i/a_i < 1/a_i 2^{n-i}.$$

In particular, the right side of $U/p - k_1/a_1 < 1/(a_1 2^{n-1})$ is very small. Hence, we can assume

$$U/p \approx k_1/a_1. \quad (4)$$

From equation (4), If we take $U' = k_1$ and $p' = a_1$, then $U'/p'$ is very close to $U/p$ and from lemma 2, the positive integers $b'_i = U'a_i \bmod p'$ for $1 \le i \le n$ are a super-increasing sequence (note that $a_i$'s are public and obvious). Subtracting the case $i = 1$ of equation (3) from the $i$-th gives

$$\frac{k_1}{a_1} - \frac{k_i}{a_i} = \frac{b_i}{a_i p} - \frac{b_1}{a_1 p} = \frac{a_1 b_i - a_i b_1}{a_1 a_i p}$$

and so, for $2 \le i \le n$,

$$|a_i k_1 - a_1 k_i| = \frac{|a_1 b_i - a_i b_1|}{p} < \frac{2p b_i}{p} = 2b_i < \frac{p}{2^{n-i-1}} \quad (5)$$

Since $a_1$ is public, It remains to compute the integer $k_1$ such that equation (5) holds, given only the integers $a_1, a_2, \ldots, a_n$. Another way to write equation (5) is

$$\left|\frac{a_i}{a_1} - \frac{k_i}{k_1}\right| < \frac{p}{a_1 k_1 2^{n-i-1}}, 2 \le i \le n. \quad (6)$$

and one sees that the problem is precisely simultaneous Diophantine approximation.

From theorem2, We can solve the simultaneous Diophantine approximation in the polynomial time and find a value for $k_1$. We now set $U' = k_1$ and $p' = a_1$ (note that $a_1$ is public) and computes $b'_i = U'a_i \bmod p'$ for $1 \le i \le n$ to obtain the sequence $(b'_1, b'_2, \ldots, b'_n)$, as we said this is a super-increasing sequence. We then compute $c' = U'c \pmod{p'}$ for any challenge ciphertext $c$. Since $(b'_1, b'_2, \ldots, b'_n)$ is super-increasing sequence, we can solve an easy knapsack problem $c' = \sum_{i=1}^{n} m_i b'_i$ with algorithm1 and therefore the original message bits $m_i, 1 \le i \le n$ are recovered.

Let $Q = 2^{n(n+1)/4} \epsilon^{-n}$ and $\lambda = \epsilon/Q$. We can use LLL-lattice basis reduction for solve equation (5) (simultaneous Diophantine approximation) and hence, the value of $U' = k_1$ is determind. Consider the lattice $L(D)$ with dimension $l + 1$ and basis matrix

$$D = \begin{pmatrix} \lambda & a_2 & a_3 & \cdots & a_l \\ 0 & -a_1 & 0 & \cdots & 0 \\ 0 & 0 & -a_1 & & \vdots \\ \vdots & \vdots & & \ddots & \vdots \\ 0 & 0 & & \cdots & -a_1 \end{pmatrix}$$

where $0 < \lambda < 1$ and $1 < l < n$. LLL-lattice basis reduction algorithm can be applied to the lattice $L(D)$ to output a relatively short vector $v = (v_1, v_2, ..., v_l)$, which can be used to approximate the simultaneous Diophantine approximation problem. Since $v \in L(D)$, there exist integers $k_1, k_2, ..., k_l$ such that

$$v = (v_1, v_2, ..., v_l)$$
$$= (\lambda k_1, k_1 a_2 - k_2 a_1, k_1 a_3 - k_3 a_1, ..., k_1 a_l - k_l a_1)$$

where $(a_1, a_2, ..., a_l)$ is the public key. After computing $v = (v_1, v_2, ..., v_l)$ with LLL-lattice basis reduction algorithm, we can compute $k_1$ from equation $v_1 = \lambda k_1$.

With the pair $(U', p') = (k_1, a_1)$, we now compute integers

$$b'_i = a_i . U' \bmod p', \quad 1 \le i \le n$$

which form lemma 2, this is a super-increasing sequence. We can use this sequence in place of to private key $(b_1, b_2, ..., b_n)$.

**Step2:**
We can eavesdrop public ciphertext $(C, D')$ from insecure channel and hence we can compute factorial carry value $U = \{u_1, u_2, ..., u_{170}\}$ of $D'$ where
$$D' = u_1 \times 169! + u_2 \times 168! + \cdots + u_{170} \times 0!$$
with the following algorithm.

---
**Algorithm4:** compute the factorial carry value of integer $m$

INPUT: integers $m, n$ such that $m < n!$.

OUTPUT: $\{u_1, u_2, ..., u_n\}$ such that $0 \le u_i \le n - i$ and $m = \sum_{i=1}^{n} u_i \times (n-i)!$.

1.     $r_0 \leftarrow m$
2.     for $i = 1$ to $n$

        do $\begin{cases} u_i \leftarrow \left\lfloor \frac{r_{i-1}}{(n+1-i)!} \right\rfloor \\ r_i \leftarrow r_{i-1} - u_i \times (n+1-i)! \end{cases}$

3.     Return "$\{u_1, u_2, ..., u_n\}$"

---

**Step3:**
We divide super-increasing sequence $(b'_1, b'_2, ..., b'_n)$ (which is computed in step1) into 8 subset public key vectors:

$$B' = \{(b'_1, b'_2, ..., b'_{170}),$$
$$(b'_{171}, b'_2, ..., b'_{340}),$$
$$\vdots$$
$$(b'_{1191}, b'_2, ..., b'_n)\}.$$

and recombine each subset public key vector using factorial carry value $U = (u_1, u_2, ..., u_{170})$ (which is computed in step2) by means of the permutation combination algorithm. Similar Hwang *et al.*'s cryptosystem, we can choose first 128 elements in each subset public key vector. Then, we will obtain 1024 elements $(b'u_1, b'u_2, ..., b'u_{1024})$.

With computed pair $(U', p')$ from step1, first compute $C' = C \times U' \bmod p'$ and then divide $C'$ into $C' = \{C'_1, ..., C'_j\}$. Each $C_k, 1 \le k \le j$ is a 1024-bit message. Now, since $(b'u_1, b'u_2, ..., b'u_{1024})$ is super-increasing sequence, we can use algorithm1 for solve the following super-increasing subset sum problems:

$$C'_1 = \sum_{i=1}^{1024} b'u_i \times x_{1,i} \bmod p'$$
$$C'_2 = \sum_{i=1}^{1024} b'u_i \times x_{2,i} \bmod p' \quad (7)$$
$$\vdots$$
$$C'_j = \sum_{i=1}^{1024} b'u_i \times x_{j,i} \bmod p'$$

and recover message bits $M_k = \{x_{k,1}, ..., x_{k,1024}\}, 1 \le k \le j$ to obtain the original message $M = \{M_1, M_2, ..., M_j\}$ for any challenge ciphertext $(C, D')$.

## 5. Performance Analysis of Attack

As we see in section 4, in step1, we need to find a pair of integers $(U', p')$ such that $U'/p'$ is very close to $U/p$ (where $U = w^{-1} \pmod p$ and $p$ are parts of the private key and $(a_1, a_2, ..., a_n)$ is public key). As we showed, we can take $p' = a_1$ where $a_1$ is public. So we need to find a value for $U' = k_1$. In step2, we can use algorithm4 for find the factorial carry value $U = \{u_1, u_2, ..., u_{170}\}$ of the public value $D'$. In step3, we need one modular multiplication for compute $C'$ and $(1024 \times j)$ subtraction to solve equation (7) with algorithm1 for recover the original message $M = \{M_1, M_2, ..., M_j\}$ from any challenge ciphertext $(C, D')$.

Hence, the more difficult and important part of attack is step1. In step1, we used simultaneous Diophantine approximation problem for finding the value of $U' = k_1$. So we need to compute the computational complexity of simultaneous Diophantine approximation problem. complexity of this problem is given in theorem3 where take $O(n^6 \max\{n \log(X), n^2 + n \log(1/\epsilon)\}^3)$ bit operations, which is polynomial time. Consequently, the proposed attack is polynomial time and practicable.

## 6. Conclusion

We considered cryptanalysis of a knapsack-type public key cryptosystem. This cryptosystem uses a combination permutation algorithm in the encryption phase to avoid the low density attack by keeping the density high. This cryptosystem is vulnerable to LLL-lattice basis reduction algorithm, since it uses a super-increasing sequence as a private key and attempt to hide this sequence with modular multiplication for constructing the public key. But as we showed, the modular multiplication cannot hide the super-increasing sequence. To avoid this attack we can choose another easy knapsack problem that is not a super-increasing sequence or we do not use modular multiplication for producing the public key from the private key.

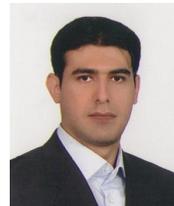

**Roohallah Rastaghi** has received his BSc degree in electrical engineering and MSc degrees in secure communication from the Aeronautical University of since and Technology, Iran in 2003 and 2010, respectively. His research interests include cryptology especially design and analyze of public-key cryptography.

**Hamid Reza Dalili Oskouei** received his BSc and MSc degrees in electrical engineering from the University of Aeronautical Science & Technology and the Trabiat Modares University, Iran in 2002 and 2004, respectively. He obtained his PhD degree in electrical engineering from Trabiat Modares University, Tehran, Iran. He then joined the University of Aeronautical Science & Technology, Tehran, Iran, as an assistant professor in 2006, his research areas are Communication, Radar,Microwave component, Antenna and wave propagation. Dr. Oskouei has served as a reviewer for a number of journals and conferences.